\magnification=1200
\nopagenumbers
\parskip=10pt
\vskip .7truein
\centerline {\bf{Comparison of Exact and Perturbative Results for
Two Metrics}}
\vskip 1.0truein
\centerline { M.Horta\c csu* , R.Kaya *$\dagger$ }
\vskip .3truein
\centerline {* Physics Department, Faculty of Science and Letters}
\vskip .1truein
\centerline{I.T.U., 80626, Maslak, Istanbul, Turkey}
\vskip .3truein
\centerline {$\dagger$ Feza G\" ursey Institute, P.K. 6, 81220,
\c Cengelk\" oy, Istanbul, Turkey}
\vskip 1.4truein

\noindent
Abstract: We compare the exact and perturbative results in two metrics
and show that the spurious effects due to the perturbation method
do not survive for physically relevant quantities such as the vacuum
expectation value of the stress-energy tensor.
\baselineskip=18pt
\footline={\centerline{\folio}}
\pageno=1
\vfill\eject
\noindent
INTRODUCTION

\noindent
Most of the effort in theoretical particle physics is on having a
consistent quantum mechanical theory which unifies the four known forces and
accounts for  all the known particles and the existing phenomenology.
A vital point in this effort is  incorporating the gravitational
interactions to the already unified scheme of the strong, electromagnetic
and weak forces .   The last resort in this heroic endeavor seems to be the
M theory $^{/1}$,
a theory in eleven dimensions which will reduce to the five consistent
string theories at the appropriate limits.

A more modest attempt is using semi-classical methods to study gravitation.
Such methods are very useful for extracting information about the theory
in the absence of a full quantization.  For instance , we can calculate
the fluctuations in the energy of a particle that propagates through
universes described by different metrics, which are exact solutions of
Einstein's equations.  Extensive work  was done in the seventies in this
field stressing the phenomena of particle production in these metrics.
This work is described in the books written by Birrell and Davis $^{/2}$,
Fulling $^{/3}$ and Wald $^{/4}$.  Here we are essentially confronted
with a problem of a particle in an external potential.  The 2n-point
functions reduce to the study of n two-point functions as
clearly described in the work by Kuo and Ford $^{/5}$.

We had applied  these methods to the calculation of the vacuum expectation
value , hereafter VEV, of the stress-energy tensor for impulsive spherical
and shock wave solutions of Nutku and Penrose $^{/6}$ and Nutku $^{/7}$
respectively.  We found a roundabout way $^{/8}$ which results in a finite
expression for the  impulsive spherical wave.  This method consisted of
taking a detour in de Sitter space for regularizing the ultraviolet 
divergences and landing in the Minkowski space after an appropriate
limit is taken.  This method seems does not seem to be able to
produce a finite result
$^{/9}$ for the shock wave $^{/7}$, though.  

We had doubts on whether the method actually gives the right answer.
We may be taking a singular limit and changing the basic character
of the problem in doing so.  Another defect may be the use of perturbation
theory which may give a different result compared to the exact one.
In  our work we stop at the second order and apply our regularization
procedure to each order seperately  taking the first finite contribution .
We thought it is worthwhile to apply our
method to a well-known case.

A second point was the presence of spurious infrared and
ultraviolet divergences
in the perturbation series and the loss of the Hadamard behaviour.
Our previous experience
$^{/9 }$ shows that this spurious
behaviour has no effect on physical quantities like the VEV of the
stress-energy tensor. Here treating two metrics both exactly and
perturbatively,
we check if such spurious behaviour exists in the perturbative approach
and if it exists, whether it is carried
to the physically relevant quantities.  Studying the same model exactly we
know that this spurious behaviour has no validity.

Work of Deser $^{/10}$   and Gibbons $^{/11}$
prohibits the existence of vacuum fluctuations for the plane wave
metrics.  Here we will first study an impulsive plane wave $^{/12}$.
We  solve the problem both exactly and  perturbatively and compare
the results.  We will show that the application of our method to the
perturbative case does not give us a result which is in contradiction with 
the exact case.  We , then apply the same method
to sandwich waves $^{/13}$.
We again solve the problem exactly , and then carry the calculation to second
order and show that there is no way of
extraction a finite expression for the VEV of the stress-energy tensor,
even if we take a detour in de Sitter space.  We end with a few remarks.

\bigskip
\noindent
{\bf{
I.  Plane  Impulsive Wave}}

\noindent
{\it{ Exact Calculation}}

\noindent
Here we take the metric describing an impulsive plane
wave $^{/12}$    ,
$$ds^2=2dudv-|d {\overline {\zeta}} + q_{\zeta \zeta} v \Theta(v) d\zeta|^2 .
\eqno {1} $$
If we take $ q=g{{\zeta^2}\over {2}}$ we get a plane wave $^{/12}$.
If the power is higher than quadratic we get pp waves $^{/14 }$.
The d`Alembertian  operator in this metric is written as
$$ L= 2 \partial_{u} \partial_{v} -{{2vg^2}\over{1-v^2g^2}} \partial_{u}
-{{1}\over{(1+vg)^2}} \partial_{x}^2 -{{1}\over {(1-vg)^2}}
\partial_{y}^2 \eqno {2} $$
where we switch to real coordinates, and define $\zeta=x+iy$.
We can reduce the problem to the Sturm-Liouville type
$$ L \phi= K \phi \eqno {3} $$
and sum over the eigenvalues to obtain the Feynman propagator.
We take
$$ \phi= f(v) e^{i(k_1x+k_2y+Ru)} \eqno {4}$$ where
$$ f(v) = {{1}\over { (1-v^2g^2)^{{1}\over{2}} \sqrt{2|R|} (2\pi)^2}}
e^{i({{k_1^2}\over { 2gR(1+vg)}}
-{{k_2^2} \over {2gR(1-vg)}}-{{Kv}\over {2R}})} \eqno {5}$$

We form the Green's function using the formula
$$ G_F= \sum_{\lambda} {{\phi(x) \phi^{*} (x')} \over {\lambda}} \eqno {6}$$
where we denote the eigenmodes $k_1, k_2, R$ and $K $ by $\lambda$.
We use the Schwinger prescription to raise the eigenvalue to the exponential
${{1}\over {K}} = -i \int_0^{\infty} d\alpha e^{i\alpha K -\alpha \epsilon} $
in the limit $\epsilon$ goes to zero.  All the integrals can be performed
easily and we find
$$ G_F= -{{\Theta (v-v')  }\over { 2\pi \sigma^2}} + {{\Theta (v'-v) }
\over {2 \pi \sigma^2}} \eqno {7}$$
where
$$ \sigma^2= 2(v-v')(u-u')- (x-x')^2(1+vg)(1+v'g) -(y-y')^2(1-vg)(1-v'g)
\eqno {8}$$
and $\Theta$ is the Heavyside unit step function.
It is clear that we do not get  a finite part for the VEV of the stress-energy
tensor which is obtained from this expression by taking the appropriate
derivatives after the coincidence limit is taken.

\noindent
{\it{ Perturbative Calculation}}

\noindent
Here we will perform the same calculation perturbatively and see if there
are spurious effects due to the perturbation algorithm.
If we expand up to second order in the coupling constant $g$, we get
$$L \approx 2 \partial_{u} \partial_{v} -2vg^2 \partial_{u}
-(1+3(vg)^2)(\partial^2_{x}+\partial^2_{y}
+2vg (\partial^2_{x}-\partial^2_{y} )). \eqno {9} $$
The zeroth-order solution gives  the free case resulting
in a Green function that goes as
$$ {{1}\over {4\pi}}{{1}\over {(u-u')(v-v')-{{1}\over {2} }
[(x-x')^2+(y-y')^2]}} \eqno {10} $$
for constant $A$.  We expand the solution in powers of $g$
and take the first order solution
as $ \phi^{(1)}= f \phi^{0}$. It is straightforward to solve for $f$ and 
we get
$$ f={{(k_1^2-k_2^2)u}\over {2iR}} \left[v+{{i}\over {R}}
-{{Ku}\over {4R^2}} \right] \eqno {11} $$
For the second order solution we  take $\phi^{(2)}= \phi^{(0)} h $.
Here $ h= v^2h_1(x,y,u)+vh_2(x,y,u)+h_3(x,y,u) .$
A straightforward calculation gives us
$$ h_1={{3i}\over {2R}} (k_1^2+k_2^2) u
-{{u^2}\over {4R^2}}(k_2^2-k_1^2)^2 ,\eqno {12} $$
$$ h_2={{u}\over{R^2}}({{K}\over {2}}-3(k_1^2+k_2^2))-{{3iu^2}\over{4R^3}}
\left( (k_1^2-k_2^2)^2+K(k_1^2+k_2^2) \right)+
{{K(k_1^2-k_2^2 )^2 u^3}\over {8R^4}}, \eqno {13} $$
$$ h_3= -i{{u}\over {R^3}}({{K}\over{2}}-3(k_1^2+k_2^2))+{{u^2}\over {R^4}}
({{1}\over{8}} \left(3(k_1^2-k_2^2)^2+3K(k_1^2+k_2^2)-K^2 \right)
$$
$$+{{iu^3}\over{8R^5}} \left(2K(k_1^2-k_2^2)^2+K^2(k_1^2+k_2^2) \right)
-{{K^2(k_1^2-k_2^2)^2 u^4}\over{64R^6}}. \eqno {14} $$
Here we see that a peculiar thing happens.  When we sum over the eigenmodes,
we get the propagator where we have powers of $(u-u')$ and $m^2$ in the
denominator. To get a finite result we need a term which has only
$(u-u')^{-1}m^{-2}$ which will be regularized
in the de Sitter space and upon differentiation gives us a finite result.
This term will be multiplied by ${{\Lambda } \over {m^2}}$ which will be
finite when we take $\Lambda$ proportional to $m^2$.  This is the only
correct choice due to dimensional reasons.  Here we do not get such a term.
The closest we get is with $ (u-u')^{-2} m^{-2}$
which has one power of $(u-u')$ too many in the denominator.

If we go to de Sitter space to cancel both the ultra-violet and
infrared divergences,we have to multiply the expression for the Green's
function by $\left(1+{{\Lambda uv}\over {6}}\right)\left(1+{{\Lambda u'v'}
\over {6}} \right)$ $^{/15}$.  We expand this expression in sums and
differences of $u,u',v,v'$.  
$$(1+{{\Lambda uv}\over {6}})(1+{{\Lambda u'v'}\over {6}})= 1+ 
{{\Lambda}\over {12}}
       \left((u+u')(v+v')+(u-u')(v-v')\right)$$
       $$+{{\Lambda^2}\over {576}}
       \left( (u+u')^2(v+v')^2-(u-u')^2(v+v')^2
       -(u+u')^2(v-v')^2+(u-u')^2(v-v')^2 \right)  $$
This process reduces the ultraviolet divergence level of the
expression by  two orders at most $^{/8, 16}$. We aim to the term which is
linear in $\Lambda$ , however, We see that the finite part
of $<T_{vv}>$ goes as
$$ <T_{vv}>\propto -2 {{\Lambda^2}\over {m^2}} \Theta(v)  \eqno {15} $$
which is finite only in  de Sitter space.  One power of the curvature
cancels with the infrared parameter since we take $ \Lambda \propto m^2 $
, but the remaining power takes the contribution to zero when we go back to
Minkowski space.  Terms with $m^4$ in the denominator that will cancel this
term have divergences which are more severe than those regulated by the
factor above.

This result which is in accord with general arguments of Deser $^{/10}$ and
Gibbons $^{/11 } $   ,
is a check that our method does not contradict any known results.

One can show that this result does not change in the presence of a pp-wave
background.  Whether a wave is plane or pp type  depends only on the
form of the function $q(\zeta)$ in the metric.
The general behaviour of the expression for the vacuum expectation
value of the stress-energy tensor does not depend on the form of the
function $q$.  This form only changes an overall factor which can not decide
whether the whole expression is finite or null. The same behaviour was already
seen in the different warp functions we have used for the spherical wave.
\bigskip
\noindent
{\bf{ 2.SANDWICH WAVE} }

\bigskip
\noindent {\it{ Exact Calculation}}

\noindent
Here we use non-flat portion of the pure gravitational sandwich metric given
by Halilsoy studied in reference 13.
At this region the metric is described by the expression
$$ ds^2= 2 du dv-cosh^2 (gu)   dx^2 -cos^2(gu) dy^2 .\eqno {16} $$
We can easily form the d'Alembertian operator
$$L= 2 \partial _{u} \partial_{v}- sech^2(gu) \partial^2_{x} -
sec^2(gu)\partial^2_y +g( tanh (gu) - tan (gu)) \partial_v . \eqno {17} $$
We fourier analyze the solution in the variables $x,y,v$ since there is
translation
invariance with respect to these variables.  Since the remaining equation is
only first order in $u$, we can easily calculate the Feynman Green's
Function for this operator
$$G_F= {{g}\over {8 \pi^2}} \Theta (u-u') {{1}\over {( cosh bu' cosbu'
cosh gu cos gu)^{{1\over {2}}}}}
{{1}\over {(AB)^{{1}\over {2}}}} {{1}\over {[(v-v') -{{g(x-x')^2}\over {2A}}
-{{g(y-y')^2}\over {2B}}]}} . \eqno {18} $$
Here
$$A= tanh gu-tanh gu' , \quad  B=tan gu-tan gu'. \eqno {19} $$
In the coincidence limit both $A$ and $B$ can be written as a power
series in $u-u'$, beginning with the linear term in $u-u'$.
The Green's Function goes as 
$$G_F \approx {{1}\over{2\pi}} {{1} \over {[ 2(u-u')(v-v')D_1-(x-x')^2 D_2
-(y-y')^2 D_3]}} .\eqno {20} $$
where $$ D_1=(1+A_1 (u-u')+B_1 (u-u')^2+...) ,\eqno {21} $$
$$D_2=( 1+A_2(u-u') +B_2(u-u')^2+...), \eqno {22} $$
$$D_3=(1+A_3(u-u')+B_3(u-u')^2+...). \eqno {23} $$
Here $A_i, B_i, i=1-3 $ are functions of $u$, but not that of $u-u'$. 

If we try to extract the VEV of the stress-energy tensor out of this 
expression  we have to first regularize it and obtain the finite part  in the
coincidence limit
before we differentiate it. Before the differentiation, say , with respect to
$u$, we can take the coincidence limit in all the other variables .  Since 
a series expansion only in  $(u-u')$ and not in the other differences exist
in the final expression, we can
not get a finite term from equation 20 in this limit.
If we go to the de Sitter space, we can get rid of the singularities and
obtain a finite result.  The curvature of the de Sitter space multiplies
our expression for this case, though.  The result goes to zero as we 
take the curvature to zero in the Minkowski limit.  
We did not encounter any infrared type singularites to cancel the de Sitter
curvature term.
We find that there are no vacuum fluctuations in this case, and the behaviour 
of the exact propagator is of the Hadamard form.
\bigskip
\noindent
{\it{Perturbative Calculation}}

\noindent
For the perturbative calculation, we take $g$, the only free parameter in our
model small we expand our operator $L$, eigenfunctions and the
eigenvalues of the associated Sturm-Liouville problem in powers of $g$.

\noindent
The operator
$L$ reads
$$L \approx 2 \partial_u \partial_v -\partial^2_x -\partial^2_y+{{m^2}
\over {2}} + g^2u^2 (\partial^2_x -  \partial^2_y) $$
$$ +g^4 u^4 \left(-{{1}\over {2}} (\partial^2_x+\partial^2_y)-{{1}\over {3}}
(\partial_u \partial_v+{{2}\over {u}} \partial_v +{{m^2}\over {4}} )\right) 
.\eqno {24} $$
Here we have added a mass term that we will use as an infrared parameter
in our calculations.  The aim is to set this term equal to zero at the end
with impunity.

The zeroth order solution gives the free Green's Function , as given in
equation 10.
The first order solution is of the form $ \phi_1= f \phi_0$, where $\phi_0$
is the zeroth order contribution.
We find 
$$\phi_0= {{1}\over {(2\pi)^2}} {{1}\over {\sqrt{2|R|}}} e^{{iKv}\over {2R}} 
e^{ik_1x} e^{ik_2y} e^{iRu} .\eqno {25} $$
$f$ is found in terms of the  fourier modes $k_1,k_2,K, R $ of $\phi_0$ ,
and is given as
$$ f=(k_1^2-k_2^2) \left( {{u^2v}\over {2iR}} + {{u}\over {R^2}}
(v+{{Kv^2}\over {4R}})
+{{i}\over {2R^3}}(2v+{{Kv^2}\over {R}}+{{K^2v^3}\over {12 R^2}} \right). 
\eqno {26} $$
For the second order  we make the same kind of ansatz , $\phi_2= g \phi_0$.
Actually this ansatz is dictated by the equations for $\phi_2$.
We find that $g$ is given as a polynomial in the variable $u$,
$$g= u^4g_1+u^3g_2+u^2g_3+u g_4+g_5 \eqno {27} $$
where $g_i, i=1-5 $ are functions of $v$ and the modes $k_1,k_2, R ,K
$ of $\phi_0$.  As a typical term we give
$$g_1=-v^2{{(k_1^2-k_2^2)^2}\over { 8R^2}}+{{v}\over {4Ri}}[-(k_1^2+k_2^2)-
{{K}\over {3}}+{{m^2}\over {6}}] .\eqno {28} $$
The other have higher powers of $v$,i.e.
$$g_2={{v^3K}\over{8iR^4}}(k_1^2-k_2^2)^2+O(v^2) ,\eqno {29} $$
$$g_3={{5K^2v^4}\over {96R^6}}(k_1^2-k_2^2)^2+O(v^3) , \eqno {30} $$
$$g_4=-{{K^3v^5}\over {96iR^8}}(k_1^2-k_2^2)^2+ O(v^4) , \eqno {31} $$
$$g_5=-{{K^4v^6}\over {1152 R^{10} }} (k_1^2-k_2^2)^2 + O(v^5) .\eqno {32} $$
Here $O(v^i)$ denotes that the highest power    of $v$ is $i$  in the sequel.

The Green's Function is calculated by summing over all the eigenmodes
$k_1, k_2, K, R$.  The calculation is standard but tedious.  It is reported in
reference 17.  We just give sample expressions from the end result.
It reads
$$G_F= {{1}\over {16\pi}} \left ( {{\Gamma_1}\over {m^6}}+{{\Gamma_2}
\over {m^4}}
+{{\Gamma_3}\over {m^2}}+{\Gamma_4}\ln(S m)+{{\Gamma_5}\over {S^2}}+
{{\Gamma_6}\over {S^4}}+{{\Gamma_6}\over {S^6}}\right) , \eqno {33} $$
where
$$S^2= (u-u')(v-v')-{{1}\over {2}} [(x-x')^2+(y-y')^2].\eqno {34} $$
$\Gamma_i ,  i=1-6 $ are functions of
$v,v',(v -v'), (x-x')^2+(y-y')^2, (x-x')^2-(y-y')^2$.
$\Gamma_1$ contains terms that are as divergent as $(v-v')^{-10} $ in the
coincidence (ultraviolet) limit.  For the others the divergences are somewhat
tamer but still existing.

In our previous work $^{/8,9}$, we had terms that go as ${{1}\over
{m^2}}\Theta(v-v')$
which had just $(u-u')$ in the denominator.  Then it was possible to cancel
this divergence by multiplying by $\Lambda (u-u')(v-v') $, which even
gave us a finite expression upon differentiation with respect to $v$ and $v'$.
Here the minimum singularity goes as ${{1}\over {(v-v')^3} }$.
There is no way to
cancel the divergence by a detour in de Sitter space, with the $\Lambda$ or
with the $\Lambda$ term.

We take this fact as a blessing.  As we have shown in the previous subsection,
we can not obtain a finite expression for $<T_{\mu \nu}>$ for this metric 
performing the calculation exactly.  The perturbative calculation, although
it gives rise to spurious infrared and ultraviolet divergences in the
intermediate steps, can not be regularized and a finite expression
can not be extracted.  We interprete this fact as the absence of vacuum
fluctuations for this case. This shows that the perturbative results do
 not contradict the exact result for the
physical quantities.

{\bf{CONCLUSION}}

\noindent

If we calculate the fluctuations
for a conformal metric, fluctuations should be absent $^{/2}$.
We  first perform perturbation theory about the Minkowski space, and
our perturbations are not strong enough to overcome the restrictions
imposed by conformal symmetry.  If we go to de Sitter space, and
perform perturbation around that metric, we do not have this obstruction.  
We always find finite fluctuations in that metric.
This argument made it possible
to extract a finite expression for the VEV of the stress-energy tensor in the
sphericalimpulsive wave metric $^{/8}$.    We also note that going to
de Sitter space also tamed our ultra-violet divergences.

We can investigate if it is generically true that taking a detour in
de Sitter space cures all the
divergence problems, or if it is a cure only for one kind of metric,
the one given by Nutku and Penrose $^{/6}$.  This trick may not reliable,
afterall.
One should compare the results with the exactly solvable cases and
check that no spurious results leak in through the perturbative method
 and the limits we used.
Here we perform the calculation both perturbatively and exactly for two
cases , and
show that there is no contradiction as far as the value for $<T_{\mu \nu}>$
is concerned.

In the spherical impulsive wave calculation, there were no dimensional
coupling constants.
It turns out that if we have dimensional coupling constants, we have
more severe ultra-violet divergences which are tamed only with having
higher powers of the curvature scalar of de Sitter space, multiplying
our expressions for the fluctuations.  This happens in the two 
metrics, plane and  sandwich waves, we have studied here .
Either we do not have severe enough  infrared divergences which
will be cancelled by $\Lambda$ or we do not have sufficient powers of
the scalar curvature term $\Lambda$
to cancel the existing  infrared divergences while its companion, powers of
$(v-v')$   is cancelling the ultraviolet ones resulting in a finite result.

\bigskip
\noindent
{\bf{
Ackowledgement:}}
\noindent
We thank Prof.Dr.Y. Nutku for his support and giving his
metrics prior to publication and for many discussions.
We thank Prof. A.N. Aliev
for illuminating discussions.  This work is
partially supported by T\" UBITAK, the Scientific and Technical Research
Council of Turkey and  M.H.'s work also by T\" UBA,
the Turkish Academy of Sciences.

\vfill\eject
\noindent
REFERENCES
\item {1.}  One may look any of the reviews present.  A manuscript
that appeared  recently depends on the plenary talk given by Gibbons
at the  GR15 meeting at Poona, India:
G.W.Gibbons, " Quantum Gravity/String/M-theory as we approach the
3rd Millennium", Cambridge Univ. preprint, gr-qc/9803065 18 Mar 1998;

\item {2.} N.D.Birrell and P.C.W. Davies, {\it{ Quantum Fields in
Curved Space}}, Cambridge University Press, Cambridge 1982;

\item {3.}  S.A.Fulling, {\it{ Aspects of Quantum Field Theory in
Curved Space-Time}}, Cambridge University Press, Cambridge 1989;

\item {4.}  R.M.Wald, {\it{ Quantum Field Theory in Curved Spacetime
and Black Hole Thermodynamics}}, The University of Chicago Press,
Chicago , 1994;

\item {5.} C.I.Kuo and L.H.Ford, Phys. Rev. {\bf{D47}} (1993) 4510;

\item {6.} Y. Nutku and R. Penrose, Twistor Newsletter {\bf{34}} (1992) 9;

\item {7.} Y.Nutku, Phys. Rev D {\bf{44}} (1991) 3164;

\item {8} M. Horta\c csu, Class. Quant. Grav. {\bf{13}} (1996) 2683;

\item {9.}M.Horta\c csu and K.\" Ulker, Class. Quant. Grav.  {\bf{15}}
(1998)1415;

\item {10.} S.Deser, J.Physics A, {\bf{8}} (1975), 1972;

\item {11.} G.W.Gibbons,  Comm. Math. Physics, {\bf{45}} (1975) 191;

\item {12.} Unpublished work of Y.Nutku; I am grateful to him giving me
this information last year.

\item {13.} M.Halilsoy, Class. Quant. Grav. {\bf{14}} (1997) 2231;

\item {14.} Unpublished work of Y.Nutku; I am grateful to him giving me
this information last year.

\item {15.} P.A.Hogan, Phys. Lett. A {\bf{171}} (1992) 21;

\item {16.} Y.Enginer, M.Horta\c csu and N.\" Ozdemir, Int. J. Mod. Phys. A
{\bf{13}} (1998) 1201;

\item {17.} R. Kaya, I.T.U. dissertation, in preparation.

\end